\documentstyle[12pt,amssym,dina4]{article}

\renewcommand{\[}{\begin{equation}}
\renewcommand{\]}{\end{equation}}

\newcommand{\R}{{\Bbb R}}
\newcommand{\Z}{{\Bbb Z}}
\newcommand{\RP}{ {{\rm \R P}} }
\newcommand{\setsep}{\,|\,}
\newcommand{\dq}{\dot q}
\newcommand{\dr}{{\dot r}}

\newcommand{\Eh}{{\cal E}_h}
\newcommand{\Qh}{Q_h}
\renewcommand{\o}{d}
\newcommand{\iso}{\simeq}
\newcommand{\Sh}{\Sigma_h}
\newcommand{\Bo}{{\cal B}}
\newcommand{\pEh}{{\partial \Eh}}
\newcommand{\id}{id}
\newcommand{\rox}{\hfill$\Box$}
\newcommand{\sm}{\setminus}
\newcommand{\ra}{\rightarrow}
\newcommand{\gb}{{b}}
\newcommand{\SP}[2]{\langle #1, #2 \rangle}
\newtheorem{proposition}{Proposition}

\begin{document}
\title{
Topology of energy surfaces and \\
existence of transversal Poincar\'e sections
}
\author{Alexey Bolsinov$^a$ \hspace{1.5em} 
        Holger R. Dullin$^b$ \hspace{1.5em} 
        Andreas Wittek$^b$ \\
\\
$^a)$ Department of Mechanics and Mathematics \\
Moscow State University \\
Moscow 119899, Russia \\
\\
$^b)$ Institut f\"ur Theoretische Physik \\
Universit\"at Bremen \\
Postfach 330440 \\
28344 Bremen, Germany \\
Email: hdullin@physik.uni-bremen.de
}
\date{February 1996}
\maketitle

\begin{abstract}

Two questions on the topology of compact energy surfaces of natural
two degrees of freedom Hamiltonian systems in a magnetic field are discussed.
We show that the topology of this 3-manifold (if it is not a unit 
tangent bundle) is uniquely determined by the Euler characteristic of the 
accessible region in configuration space. In this class of 3-manifolds
for most cases there does not exist a transverse and complete Poincar\'e
section. We show that there are topological obstacles for its existence
such that only in the cases of $S^1\times S^2$ and $T^3$ such a 
Poincar\'e section can exist.

\end{abstract}

\section{Introduction}

The question of the topology of the energy surface of Hamiltonian
systems was already treated in the 20's by Birkhoff and
Hotelling \cite{Hotelling25,Hotelling26}. 
Birkhoff proposed the ``streamline analogy'' \cite{Birkhoff22}, 
i.e.\ the idea that the flow of a  Hamiltonian system on the 3-manifold could 
be viewed as the streamlines of an incompressible fluid evolving in this
manifold. Extending the work of Poincar\'e \cite{Poincare1892} he noted that it
might be difficult to find a transverse Poincar\'e section which is 
complete (i.e.\ for which every streamline starting from the surface of section
returns to it) \cite{Birkhoff17}. Hotelling classified some of the topologies of energy 
surfaces with two degrees of freedom. 1970 Smale initiated 
the study of ``Topology and Mechanics'' \cite{Smale70} 
from the modern point of view.
This work had a great influence and stimulated a lot of research 
especially in the Russian school of mathematics, 
see e.g.~\cite{Fomenko91,Kharlamov83,Oshemkov91,Bolsinov91,Tatarinov73,Iacob71}.

We want to take the present knowledge about the topology of energy
surfaces of natural Hamiltonian systems and return to the question of
Birkhoff about the existence of transverse and complete Poincar\'e surfaces
of section. The list of topologies of natural Hamiltonian systems is in
principle known, but here we collect the
results we need and give a proof using Heegard splittings
which explicitly constructs an embedding of the split halves of our
``manifold of streamlines'' into $\R^3$. 
With the help of the computer it is possible to create a realistic
picture of Birkhoffs ``streamline analogy'' using our result.
In the second part the list of topologies of energy surfaces
is compared 
to the list of manifolds that can have a complete and transverse 
Poincar\'e section, i.e.\ which admit the structure of a bundle over $S^1$ 
with a Riemann surface as a fiber. In \cite{DW95} we already noted
that there can be topological obstacles for the existence of a
transverse and complete Poincar\'e section. 
We now show that in the class of all energy surfaces of natural
Hamiltonian systems (possibly with a magnetic field) there can
only exist a transverse and complete Poincar\'e section if the 3-manifold
is a direct product of $S^2$ or $T^2$ with $S^1$.

\section{Topology of Energy Surfaces}

Consider a natural time independent Hamiltonian system with two degrees of 
freedom in a magnetic field. 
The smooth and orientable two dimensional configuration space is denoted by 
$Q$. The system is described by the Lagrangian on the tangent bundle $TQ$ 
given by
\[ \label{eqn:lag}
        L(q,\dq) = \frac{1}{2} \SP{\dq}{ T(q) \dq} - V(q) + \SP{A(q)}{\dq}, 
\]
with a positive definite matrix $T(q)$, potential $V(q)$ and
vector potential $A(q)$. 
Since $\det T \not = 0$ the momenta are $p=\partial L/\partial \dq$ 
and the Legendre transformation to $T^\ast Q$ gives the Hamiltonian 
\[
        H(q,p) = \frac{1}{2} \SP{(p-A(q))}{ T^{-1}(q) (p-A(q)} + V(q).
\]
If $Q$ is compact it is a Riemann surface $R^2_g$ whose genus we denote by $g$,
otherwise $Q$ is the Euclidean plane $\R^2$ or a cylinder $\R^1\times S^1$.
The accessible region $\Qh$ in $Q$ for fixed energy $H=h$ is 
the set of points in $Q$ for which the potential energy does not
exceed the total energy
\[
        \Qh = \{ q \in Q \setsep V(q) \le h \},
\]
which we assume to be compact.
Each connected component of $\Qh$ can be treated separately.
The ovals of zero velocity with $\dq = 0$ or equivalently $V(q)=h$ 
are the boundaries of $\Qh$, if any. The number of ovals 
of zero velocity, i.e.\ the number of disjoint components of 
$\partial \Qh$ is denoted by $\o$.
By abuse of language we denote the parts of $Q$ which are excluded
from $\Qh$ by the ovals of zero velocity as ``holes'' in $Q$.
The energy surface
\[ \label{eqn:Eh}
        \Eh = \{ (q,p) \in T^\ast Q \setsep H(q,p) = h \}
\]
is compact because $\Qh$ is assumed to be compact. 
By this assumption we have $\o > 0$ for $Q = \R^2$ 
and $\o>1$ for $Q = \R\times S^1$.
In the following we will include the cases of non-compact
$Q$ into the case of $Q \iso S^2$ because a disc with $\o-1$ holes
(not counting the outer boundary of the disc)
is homeomorphic to a sphere with $\o$ holes, similarly for a cylinder
with $\o-2$ holes.
Therefore the topology of $\Eh$ only depends on the genus $g$ of $Q$ and
on the number of ovals of zero velocity $\o$ of $\Qh$.
Note that $\Qh$ is the projection of $\Eh$ onto $Q$.

The case of $\o=0$, i.e.\ the motion on a compact Riemann surface $Q=R^2_g$
(with sufficiently high energy $h > V(q)$ everywhere) almost by definition 
(\ref{eqn:Eh}) has an energy surface homeomorphic to the unit tangent bundle 
of $R^2_g$. Here we want to classify all the other cases with $\o>0$.

\begin{proposition}
The topology of the energy surface $\Eh$ of a two degree of freedom
Hamiltonian system is determined by the Euler characteristic $\chi$ 
of the accessible region of configuration space $\Qh$ 
if there is at least one oval of zero velocity.
\end{proposition}

Our proof is elementary and constructive: We embed $Q$ in $\R^3$ and
attach ellipses of possible velocity to every point of $\Qh$.
Cutting these velocity ellipses we obtain a Heegard splitting of $\Eh$ from
which the topology of $\Eh$ is determined.

Since $Q$ is an orientable Riemann surface it can be embedded in $\R^3$:
\[
        Q \iso \{ r \in \R^3 \setsep F(r) = 0 \}.
\]
In the Lagrangian (\ref{eqn:lag}) we now choose $r$ as global coordinates 
with the additional constraint $F(r)=0$.
The energy function $\tilde E(q,\dq)$ on $TQ$ is given by
\[
        \tilde E(q,\dq) = \frac{1}{2} \SP{\dq}{ T(q) \dq} + V(q),
\]
and similarly 
\[
        E(r,\dr) = \frac{1}{2}  \SP{\dr}{ \tilde T(r) \dr} + 
                \tilde V(r), \qquad \SP{F_r}{\dr} = 0, 
\]
where $\tilde T|_Q = T(q)$ and $\tilde V|_Q=V(q)$ and the tildes are
omitted in the following. The reason for treating everything on $TQ$
instead of $T^\ast Q$ is that the linear terms in the momenta in the
Hamiltonian due to the vector potential $A$ are not present
if the energy is treated as a function of the velocities $\dq$.
Moreover, note that with non-vanishing $A$ 
on the boundary of $\Qh$ we have zero velocity $\dq$ but not zero momentum $p$.

With $E(r,\dr)$ we have an embedding of $\Eh$ into Euclidean space $\R^6$ 
given by
\[
        \Eh \iso \{ (r,\dr) \in \R^6 \setsep E(r,\dr)=h, F(r)=0, 
                \SP{F_r}{\dr} = 0 \}. 
\]

Following Birkhoff, Hotelling and Smale 
\cite{Birkhoff27,Hotelling25,Hotelling26,Smale70} the energy surface 
is constructed by attaching circles in velocity space to every point in the 
(accessible) configuration space $\Qh$. This gives a fiber bundle with 
base $\Qh$ and fiber $S^1$ where the fiber is contracted to a point on
$\partial\Qh$.
In our embedding this means to take 
any point $r$ on $Q \subset \R^3$ and to calculate the remaining kinetic energy
$h-V(r)$. Outside $\Qh$ it is negative, on the boundary it is zero and inside
of $\Qh$ it is positive. In the latter case the possible
velocities are given by $\SP{\dr}{ T \dr} = 2(h-V(r))$. The constraint ensures
that $\dr$ is in the tangent plane of $F(r)=0$ at $r$. Therefore the possible
velocities are located on an ellipse in the tangent plane.

In order to cut the velocity ellipses at every point we need a device
to fix a zero position on this $S^1$, i.e.\ we want to construct a
global section for the fiber bundle. This global section can be constructed
with the help of a nowhere vanishing vector field $\xi$ on $\Qh$.
On a Riemann surface $Q$ of genus $g \not = 1$ there does not exist a
vector field $\xi$ without equilibrium points. If, however, there are
holes (or punctures) in the Riemann surface we can construct $\xi$ on it,
such that the restriction
to $\Qh$ is without singularities, essentially by moving the singularities
into the hole(s). Note that at this point the assumption $\o > 0$ 
is necessary (except for the case of $Q=T^2$).
Let $\xi(r)$ be specified in the embedding in $\R^3$
such that $\SP{\xi(r)}{F_r} = 0$. 
Denote by $N(r)$ the normal vector of the surface $F(r)=0$. 
Using $\xi(r)$ every nonzero velocity 
ellipse can be cut into two halves specified by 
$\SP{N(r)}{\xi(r) \times \dr} \ge 0$
and $\SP{N(r)}{\xi(r) \times \dr} \le 0$, 
the two halves joining at the place where
$\xi$ and $\dr$ are (anti)-parallel.
In this way we cut $\Eh$ into two 
topological equivalent pieces
\[
        \Eh^\pm = \{ (r,\dr) \in \Eh \setsep \pm
        \SP{N(r)}{\xi(r)\times \dr} \ge 0 \} 
\]
such that 
\[
        \Eh = \Eh^+ \cup \Eh^- \quad\mbox{and}\quad 
        \pEh^+ = \pEh^- = \Eh^+ \cap \Eh^-
\]
The two pieces can be embedded into $\R^3$ in the following way.
Each half of the velocity ellipse is parametrized by the 
scalar product $\SP{\xi(r)}{\dr}$. The embedding is defined by
\[ 
\begin{array}{rcl}
        M : \Eh^\pm & \rightarrow & \R^3 \\
                (r,\dr) & \mapsto & r + \alpha N(r) \SP{\xi(r)}{\dr}, 
\end{array}     
\]
where $\alpha$ is a sufficiently small constant in order for $M$ to be
a homeomorphism. 
The two solid handle-bodies $M(\Eh^+)$ and $M(\Eh^-)$ coincide 
in $\R^3$ and define a Heegard splitting of $\Eh$
\cite{SeifertThrelfall80,Zieshang94,DFN92}: 
Their boundaries $M(\pEh^\pm)$ have to be identified to reobtain $\Eh$. 
The most important point is that
the gluing homeomorphism from $M(\pEh^+)$ to
$M(\pEh^-)$ is the identity map, as is obvious from our construction. 
The topology of $\Eh$ is therefore
completely determined by the topology of $M(\Eh^\pm)$, which in turn
is determined by its boundary $\partial M(\Eh^\pm) = M(\pEh^\pm) = \Bo$.

The solid handle-body $M(\Eh^\pm)$ can be thought of as a ``thickened'' $\Qh$ 
because it is obtained by attaching small intervals in the direction of the 
normal to every interior point, while the interval is contracted to a point
on the ovals of zero velocity $\partial\Qh$.
The boundary $\Bo$ of the solid handle-body is obtained by deleting all the 
interior points of the attached intervals. The resulting Riemann 
surface is obtained from two copies of $\Qh$ (corresponding to the two
endpoints of each interval) glued together along the ovals of
zero velocity $\partial\Qh$. Analogous to the construction of the
energy surface as a bundle over $\Qh$ with fiber $S^1$ we can think
of $\Bo$ as a bundle over $\Qh$ with fiber $S^0$ (i.e.\ two points)
where the two points are identified on $\partial\Qh$.

$\Qh$ is determined by the genus of $Q$ and the number of holes $\o$.
The Euler characteristic $\chi$ of $\Qh$ is $2-2g-\o$ because every
hole removes one triangle from the triangulation of $R^2_g$ which decreases 
$\chi$ by one \cite{SeifertThrelfall80}. 
To calculate $\chi(\Bo)$ we just double $\chi(\Qh)$ 
because gluing two holes (i.e.\ triangles) leaves $\chi$ unchanged:
\[
        \chi(\Bo) = 2 \chi(\Qh) = 4-4g-2\o = 2 - 2(2g+\o-1),
\]
such that we obtain $2g+\o-1 = 1-\chi(\Qh)$ for the genus of $\Bo$.
This proves that the topology of $\Eh$ is determined
by the Euler characteristic $\chi$ of $\Qh$. 
\rox \vspace{1ex}

\noindent Denote the genus of $\Bo$ by $\gb = 2g+\o-1$. 
For $\gb=0,1$ there is only one possibility for $\Qh$,
namely with $g=0$ and $\o=1,2$. 
But for larger $\gb$ we obtain a nontrivial equivalence
of energy surfaces for systems on different configuration spaces. 
The first example of nontrivial equivalence
is obtained for $g=0$, $\o=3$, i.e.\ a sphere with
three holes, which topologically gives the same energy surface 
as for $g=1$, $\o=1$, i.e.\ a torus with one hole. The former system
can be realized, e.g., by certain spinning tops while the latter occurs,
e.g., in the double pendulum \cite{DW95}. 
Note that the $\Qh$ in these examples are
not homeomorphic to each other, even though their Euler characteristic
is the same. 
Most notably for the spinning top $\Qh$ can be mapped
to $\R^2$ while for the double pendulum this is impossible.

Our next task is to show that if $\chi(\Qh)$ is different for two
energy surfaces
then they are not homeomorphic and to actually determine the topology of $\Eh$.
The result is well known, in principle, because it 
follows from the Heegard splittings obtained above, see 
e.g.~\cite{SeifertThrelfall80,Zieshang94,DFN92}: 
We nevertheless give an elementary argument for the cases we need.

\begin{proposition}
Let there be at least one oval of zero velocity in $\Qh$ and denote
the Euler characteristic by $\chi(\Qh)=1-\gb$. 
For $\gb=0$ the energy surface $\Eh$ is homeomorphic to $S^3$.
For $\gb > 0$ $\Eh$ is homeomorphic to the connected sum of
$\gb$ copies of $S^1 \times S^2$.
\end{proposition}

We choose the simplest case $g=0$ which allows for all possible $\gb$.
Since $\o > 0$ we can map the accessible region $\Qh$ of the sphere $S^2$
to the Euclidean plane. The resulting disc $D^2$ has $\gb=\o-1$ holes.
An example can be constructed by considering $H = p^2/2 + V(r)$
($z=0$, $p_z=0$),
with $\gb$ distinct points $r_i$ and $V(r) = r^2 + \sum 1/|r-r_i|$

For all $\gb$ we have $N(r) = (0,0,1)$ and can, e.g., take $\xi(r) = (0,1,0)$ 
as a vector field, such that $\SP{\xi}{\dr} = \dot y$.
For this special choice of $\xi$ we actually think of $M(\Eh^\pm)$
as a projection of $\Eh$ into the Euclidean space $(x,y,\dot y)$,
which produces a double cover in the interior because the sign of 
$\dot x$ is lost.

For $\gb=0$ $\Qh$ is a disc $D^2$ 
and attaching the intervals of allowed $\dot y$ at fixed $r$
(corresponding to each half of the ellipse of possible velocities)
we obtain a ball $D^3 \iso M(\Eh^\pm)$. 
Gluing two $D^3$ along their common boundary $S^2 \iso M(\pEh^\pm)$ 
gives $S^3 \iso \Eh$. 

If $\gb=1$ then $\Qh$ is an annulus. 
Attaching the intervals of $\dot y$ we now obtain a solid torus
$D^2\times S^1$, whose boundary is $T^2$, i.e.\ $\gb=1$.
Gluing two solid tori by the trivial identification along their
boundary gives $S^1\times S^2$. 
Note that the trivial identification is important for $\gb>0$, because taking 
a different gluing homeomorphism would yield a different 3-manifold,
e.g., $S^3$ or $\RP^3$ for $\gb=1$. 

In the case of $\gb=2$ we have $\Qh$
homeomorphic to a disc $D^2$ with two holes inside. Attaching the
intervals we obtain a solid handle-body $M(\Eh^\pm)$ whose 
boundary $\Bo$ has genus $\gb=2$. Now we cut
this handle-body into two parts separating the two holes, such that
we obtain two solid tori $S^1\times D^2$. The cut is along a disc $D^2$.
Gluing each solid torus to its partner (leaving the $D^2$ of the cut
unidentified) we obtain $S^1\times S^2$ with a solid ball $D^3$ removed.
The boundary of this $D^3$ is $S^2$, which is obtained by gluing the 
$D^2$ of the cut to its partner along their boundary. Now we have to 
restore the cut to obtain $\Eh$, i.e., we have to glue two 
copies of $S^1\times S^2$ along the boundary of a $D^3$ removed from 
the two copies. This is exactly the operation of the connected sum and
we obtain $\Eh \iso S^1\times S^2 \# S^1\times S^2$
(sometimes this manifold is denoted by $K^3$).
For $\gb > 2$ the same process is repeated $\gb$ times and we obtain
the connected sum of $\gb$ copies of $S^1\times S^2$. 
\rox \vspace{1ex}

\noindent We summarize our results in the following table, where $M\sm nD^2$
denotes the two dimensional manifold $M$ with $n$ disks $D^2$ removed.
\begin{table}
$$
\begin{array}{c|r|ccccl}
&\chi(\Qh) &  1  & 0             &  -1   & \cdots & 1-\gb \\ 
&\gb    &  0    & 1             &  2    & \cdots & \gb \\ 
\hline
&\Eh    & S^3   & S^1\times S^2 & 
                                  \#^2 S^1 \times S^2 
                                        & \cdots & \#^\gb S^1\times S^2 \\ 
 &\Bo   & S^2   & T^2           & R^2_2 & \cdots & R^2_\gb \\ 
\hline
\hline
Q & \Qh & & & & & \\ 
\hline
S^2  &  & S^2\sm D^2   & S^2\sm 2D^2   & S^2 \sm 3D^2  & \cdots & S^2 \sm (\gb+1) D^2 \\
\R^2 &  & D^2           & D^2\sm D^2    & D^2\sm 2D^2   & \cdots & D^2 \sm \gb D^2 \\
\R\times S^1& &         & D^1 \times S^1 & D^1\times S^1\sm D^2 &\cdots&D^1\times S^1\sm (\gb-1)D^2 \\
T^2  &  &               &               & T^2 \sm D^2   & \cdots & T^2 \sm (\gb-1) D^2 \\
\vdots& &               &               &               &       & \vdots \\
R^2_g&  &               &               &               &       & R^2_g \sm (\gb+1-2g) D^2 \\
 &      &               &               &               &       & \vdots \\
 &      &               &               &               &       & R^2_{[\gb/2]} \sm D^2
\end{array}
$$
\end{table}
By a recent result from Kozlov and Ten \cite{KozlovTen96}
all of the combinations of $\Qh$ and $\Eh$ listed in the table 
can even be realized by natural 
Hamiltonian systems that are completely integrable.

Let us remark that in the cases without ovals of zero velocity the
energy surface is the unit tangent bundle. Most notably 
$Q=\Qh \iso S^2$ gives $\Eh \iso \RP^3$ and $Q=\Qh\iso T^2$ gives 
$\Eh\iso T^3$.
All possible $\Eh$ can be realized by mechanical systems. However, most
often one encounters $S^3$, $S^1\times S^2$, $\RP^3$ and $T^3$.
In the dynamics of the spinning top one can find $\#^2S^1\times S^2$ 
and $\#^3 S^1 \times S^2$
\cite{Iacob71,Tatarinov73,Oshemkov91}.

\section{Nonexistence of complete transverse Poincar\'e sections}

With the complete list of topologies of compact energy surfaces of natural
Hamiltonian systems with two degrees of freedom at hand we now want
to show that in all cases except $S^1 \times S^2$ and $T^3$ a complete
transverse Poincar\'e section is impossible. 
Using the result of the last section we see that the two exceptions are
obtained from $\Qh$ that contain $S^1$ as a trivial factor, i.e., for
$\Qh \iso S^1\times D^1$ or $T^2$.

Let the Poincar\`e section be defined by a smooth function
$S(q,p)=0$ on phase space.
The surface of section $\Sh$ is obtained by restriction to
the energy surface,
\[
        \Sh = \{ (q,p) \in T^\ast Q \setsep H(q,p)=h, S(q,p)=0 \}.
\]
If there are more than one component each of them can be treated
separately.
Excluding cases with critical points, $\Sh$ is a Riemann surface
of arbitrary genus embedded in $\Eh$. 
The equations of motion are given by the Poisson bracket
$\dot F = \{F,H\}$.
The surface of section is transversal to the flow if
\[
        \dot S|_{(q,p)} \not = 0 \quad\mbox{for all}\quad (q,p) \in \Sh.
\]

Then the Poincar\`e map $P : \Sh \rightarrow \Sh$ is defined by
\[
        (q,p) \in \Sh \mapsto g^{\tau(q,p)} (q,p) \in \Sh 
\]
where $g^t$ denotes the Hamiltonian flow and $\tau(q,p)$ is the first return 
time. 
We assume that the section is transverse and $\Sigma$-complete 
\cite{DW95}, i.e., that every orbit starting on $\Sh$ returns to $\Sh$ and
therefore $\tau$ is finite and $P$ is well defined on all of $\Sh$.
A Poincar\'e section with these properties will be called complete and
transverse in the following.
The Poincar\'e map 
$P$ has degree one due to the existence and uniqueness of the solutions 
of the differential equation which connects preimage and image by an
integral curve. Hence $\Eh$ has the structure of a fiber bundle with
base $S^1$ and fiber $\Sh$ \cite{DFN92}. Let $\phi$ be in $S^1$.
For every base point $\phi$ the fiber consists of all points
$g^{\phi\tau(q,p)/2\pi}(q,p)$ such that $(q,p)\in\Sh$ for $\phi=0$.
The converse formulation is that given the Poincar\'e mapping the
energy surface can be obtained by a suspension 
into a flow which automatically creates the structure of fiber bundle with
base $S^1$ and fiber $\Sh$.

In \cite{DW95} we have shown that for $\Eh \iso S^3$ the existence of
a transverse section is in contradiction
with Liouville's preservation of phase space volume. In the following we
use different arguments based on the above bundle structure to treat 
the general case.

\begin{proposition}
A complete transverse Poincar\'e section for a natural two degree of freedom
Hamiltonian system can only exist for energy surfaces homeomorphic
to $S^1\times S^2$ or $T^3$. If it exists it can only be realized by
the trivial bundle.
\end{proposition}

Let us assume there exists a complete transverse section in $\Eh$.
This implies that $\Eh$ admits the structure of a fiber bundle 
with base $S^1$ and fiber $\Sh$
\[ \label{eqn:bundle1}
        \Eh \stackrel{\Sh}{\longrightarrow} S^1,
\]
as already explained. 
Let us consider the exact homotopy sequence of this bundle:
\[
\begin{array}{ccccccc}
\pi_2(S^1) & \ra & \pi_1(\Sh) & \ra & \pi_1(\Eh) & \ra & \pi_1(S^1) \\
  0        & \ra & \pi_1(\Sh) & \ra & \pi_1(\Eh) & \ra & \Z
\end{array}
\]
which implies 
\[ \label{eqn:facgroup}
        \pi_1(\Eh) / \pi_1(\Sh) = \Z \quad
\]
and therefore
\[ \label{eqn:subgroup}
                \pi_1(\Sh) \subset \pi_1(\Eh).
\]
If $\Eh$ is a direct product with $S^1$ this is obviously possible
because $\pi_1(M\times N) = \pi_1(M)\times \pi_1(N)$.
For the energy surface $S^1\times S^2$ and $T^3$ we have the trivial
bundles as a possible solution.

For all other energy surfaces of natural system the bundle structure
(\ref{eqn:bundle1}) is impossible.
We first treat the cases with ovals of zero velocities.
Because $\pi_1(S^3)=\id$ (\ref{eqn:subgroup}) can not hold because 
$\id$ does not have a nontrivial subgroup. Therefore the energy surface $S^3$ 
does not admit the bundle structure (\ref{eqn:bundle1}) and therefore does 
not admit a complete transverse section. 
For all the other cases of energy surfaces where there are ovals
of zero velocity in $\Qh$ we have $\gb \ge 1$ and
\[ \label{eqn:pirbig}
 \pi_1(\#^\gb S^1\times S^2) = \pi_1(S^2 \times S^1) \ast \ldots 
        \ast \pi_1(S^2 \times S^1)
        = \Z \ast \Z \ast \ldots \ast \Z,
\]
i.e., $\pi_1$ is a free group with $\gb$ generators. Every subgroup
of a free group is a free group, so by (\ref{eqn:subgroup})
$\pi_1(\Sh)$ must be a free group. But $\Sh$ is a Riemann surface $R^2_s$
of arbitrary genus $s$ and $\pi_1$ of any Riemann surface never is a 
free group for $s>0$ \cite{DFN92}, which gives us a contradiction.
For $s=0$ we have $\pi_1(S^2)=\id$ and (\ref{eqn:facgroup})
gives $\pi_1(\Eh)=\Z$, which contradicts (\ref{eqn:pirbig}) 
because $\gb \ge 2$.

We now turn to energy surface obtained from $\Qh$ without ovals of zero 
velocity. If $Q=\Qh = S^2$ we have $\Eh\iso\RP^3$ and $\pi_1(\RP^3)=\Z_2$ is 
a finite group so that similar
to the case of $S^3$ equation (\ref{eqn:subgroup}) can not be fulfilled.
For $Q=\Qh=T^2$ we have already seen that $\Eh=T^3$ admits a complete
transverse section.

The remaining cases are the energy surface obtained from $Q = \Qh = R^2_g$ 
with $g>1$, i.e., the corresponding unit tangent bundles. 
These energy surfaces already carry a bundle
structure, but with base $R^2_g$ and fiber $S^1$
\[ \label{eqn:classU}
        \Eh \stackrel{S^1}{\longrightarrow} R^2_g, \quad g \ge 2
\]
as opposed to the required structure for a 
complete transverse section in (\ref{eqn:bundle1}) with $\Sh = R_s^2$,
\[ \label{eqn:classP}
        \Eh \stackrel{R^2_s}{\longrightarrow} S^1, \quad s \ge 0.
\]
Denote each of the
unit tangent bundles described by (\ref{eqn:classU}) by $U$
and each of the manifolds admitting a complete transverse Poincar\'e 
section by $P$.
We now show that $G_u = \pi_1(U)$ and $G_p = \pi_1(P)$ are different
for any choice of $U$ and $P$. 
The method of proof is inspired by \cite{Zieshang80b}.

Let us first treat the cases with $s \ge 2$.
As usual for any group $G$ denote by $C(G)$ is center and
by $[G,G]$ its commutant. Now $U$ and $P$ are different because 
\[ \label{eqn:ggcid}
        [G_p,G_p] \cap C(G_p) = \id,
\]
but
\[ \label{eqn:ggcnoid}
        [G_u,G_u] \cap C(G_u) \not = \id
\]
contains at least an infinite cyclic group.

In (\ref{eqn:facgroup}) we observed that $G_p$ contains a normal
subgroup $G'_p$ isomorphic to $\pi_1(\Sh)$, such that 
$G_p/G'_p = \Z = \pi_1(S^1)$. 
In particular the factor group is commutative.
Since the commutant is a minimal normal subgroup such that the corresponding 
factor group is commutative we have $[G_p,G_p] \in G'_p$.
But the fundamental group of a Riemann surface has no center, i.e., 
$C(G'_p)=\id$. Therefore, $G'_p$ does not intersect
with $C(G_p)$ and so does not $[G_p,G_p]$, because it is a subgroup
of $G'_p$ and we obtain (\ref{eqn:ggcid}).

Consider the group $G_u$ now (see, e.g., \cite{DFN92}). 
It can be represented as the group generated by $a_1, b_1, \ldots, a_g, b_g$ 
with the following relations:
\begin{enumerate}
\item
Let $\alpha$ be the Euler number of the unit tangent bundle $U$. Then
\[
        a_1 b_1 a_1^{-1}b_1{-1} \ldots a_g b_g a_g^{-1} b_g^{-1} = z^\alpha.
\]
In our case $\alpha$ is just the Euler characteristic $\chi = 2-2g$.
\item
$z$ commutes with any element of the group $G_u$. In particular
$z^\alpha$ belongs to $C(G_u)$. 
\end{enumerate}
But it is easily seen from the first relation
that $z^\alpha \in [G_u,G_u]$. So the intersection 
$[G_u,G_u] \cap C(G_u)$ contains a least $z^\alpha$
as stated in (\ref{eqn:ggcnoid}).
Therefore we have shown that the fundamental groups are different,
so the manifolds $U$ and $P$ are also different. 

Now we have to consider the case of $s=0,1$, i.e.\ where the surface 
of section $\Sh$ is $S^2$ or $T^2$.
In the case of the torus $\pi_1(P)$ can be
generated by three generators $a$, $b$, and $z$ with the relations
\[\label{eqn:trel}
\begin{array}{rcl}
        ab & = & ba \\
        zaz^{-1} & = & \phi(a) \\
        zbz^{-1} & = & \phi(b) 
\end{array}
\]
where $\phi$ is some automorphism of the fundamental group of $T^2$,
i.e., the commutative subgroup generated by $a$ and $b$.
It follows from this that the 1-homology group has 
at least one generator of infinite order and no more than
3 generators. But the corresponding homology group for $U$ has more
than 3 generators ($g\ge2$).
In the case of the sphere as a surface of section the first homology group
of $P$ is $\Z$
so the same argument holds as for $T^2$, which concludes the proof that
for the unit tangent bundles of $R^2_g$ with $g\ge2$ there does not
exist a complete transverse section.

Finally we show that for the cases $\Eh=S^1\times S^2$ and $\Eh=T^3$
where a complete transverse section exists it can only be constructed 
from the trivial bundle with $\Sh=S^2$ resp.\ $\Sh=T^2$.
Recall that in both cases $\pi_1(\Eh)$ is commutative.
Now both manifolds can not be realized as $S^1$ bundles with base 
$R_g^2$, $g\ge 2$, because the homotopy group of this bundle
contains the non commutative homotopy
group of the base $R_g^2$ as a subgroup, see~(\ref{eqn:subgroup}).
In the case of $\Eh\iso T^3$ we have $\pi_1(\Eh)=\Z^3$. But
as already mentioned the homology group of the $\Sh=S^2$ bundle over
$S^1$
has only one generator, so that the only possibility 
is with $\Sh=T^2$. Moreover, we must have $\phi=\id$ in (\ref{eqn:trel}) 
in order to obtain $\Z^3$.
Therefore the Poincar\'e section must be obtained
from the trivial bundle.
For $\Eh\iso S^1\times S^2$ we have $\pi_1(\Eh) = \Z$.
But the homotopy group of the $S^1$ bundle of $\Sh=T^2$ contains 
$\pi_1(\Sh) = \Z^2$ as a subgroup so 
the only possibility is with $\Sh=S^2$.
Finally the only orientable $S^2$ bundle over $S^1$ is the trivial 
bundle. 
\rox

\section{Discussion}

The difficulty in establishing a complete transverse section was already noted 
by Birkhoff \cite{Birkhoff17}, who required a coordinate
transformation to exist, which globally introduces an angle $\phi$
in $\Eh$ and moreover that $\dot \phi \not = 0$. We were not dealing with
the second requirement here. Instead we have shown that there are topological
obstacles for the existence of such an angle in most energy surfaces,
{\em independently} of the dynamics. We established the
nonexistence of complete transverse sections in most cases. For the question 
of existence in the exceptional cases $\Eh\iso S^1\times S^2$ or $T^3$ the 
dynamical system has to be considered, i.e.\ Birkhoffs second condition
has to be checked.
This additional condition makes it difficult to find complete transverse
sections even in the two special cases.
In \cite{DW95} we have
shown that for time reversal Hamiltonians transverse
sections that are time reversal symmetric are impossible. 
The only examples of complete transverse sections
(except for a trivial time periodic forcing which we are not considering here)
we know of, have a strong vector potential $A(q)$ breaking the
time reversal symmetry (of course their $\Qh$ must be 
a torus or a cylinder) \cite{DW95}.
These considerations have been our motivation to drop the 
requirement of transversality and instead 
try to construct Poincar\'e sections that are complete, see \cite{DW95}.
We suspect that it is impossible to find a transverse and complete 
Poincar\'e section for a natural time reversible Hamiltonian
system.

\section*{Acknowledgements}

We would like to thank A.T.~Fomenko, S.V.~Matveev, A.A.~Oshemkov
and P.H.~Richter for useful discussions. 
This work was partially supported by the Russian Foundation
for Fundamental Science (project 95-01-01604) and by the
Deutsche Forschungsgemeinschaft.

\bibliographystyle{plain}

\begin{thebibliography}{10}

\bibitem{Birkhoff17}
G.~D. Birkhoff.
\newblock Dynamical systems with two degrees of freedom.
\newblock {\em Trans. Am. Math. Soc.}, 18:199--300, 1917.

\bibitem{Birkhoff27}
G.~D. Birkhoff.
\newblock {\em Dynamical Systems}.
\newblock American Mathematical Society, Providence, RI, 1922.

\bibitem{Birkhoff22}
G.~D. Birkhoff.
\newblock Surface transformations and their dynamical application.
\newblock {\em Acta Math}, 43:1--119, 1922.

\bibitem{Bolsinov91}
A.~V. Bolsinov.
\newblock Methods of calculation of the fomenko-zieschang invariant.
\newblock In A.~T. Fomenko, editor, {\em Topological Classification of
  Integrable Systems}, volume~6 of {\em Adv. in Soviet Mathematics}, pages
  147--183, Providence, RI, 1991. American Mathematical Society.

\bibitem{DFN92}
B.~A. Dubrovin, A.~T. Fomenko, and S.~P. Novikov.
\newblock {\em Modern Geometry - Methods and Applications, Part III.}
\newblock Springer, Berlin, 1992.

\bibitem{DW95}
H.~R. Dullin and A.~Wittek.
\newblock Complete {P}oincar{\'e} sections and tangent sets.
\newblock {\em J.~Phys.~A}, 28:7157--7180, 1995.

\bibitem{Fomenko91}
A.~T. Fomenko.
\newblock Topological classification of all integrable {H}amiltonian
  differential equations of general type with two degrees of freedom.
\newblock In T.~Ratiu, editor, {\em The Geometry of Hamiltonian Systems},
  volume~22 of {\em Math. Sci. Research Institute Publ.}, pages 131--339, New
  York, 1991. Springer.

\bibitem{Hotelling25}
H.~Hotelling.
\newblock Three-dimensional manfiolds of states of motion.
\newblock {\em Trans. Am. Math. Soc.}, 26:329--344, 1925.

\bibitem{Hotelling26}
H.~Hotelling.
\newblock Multiple-sheeted spaces and manfiolds of states of motion.
\newblock {\em Trans. Am. Math. Soc.}, 27:479--490, 1926.

\bibitem{Iacob71}
Andrei Iacob.
\newblock Invariant manifolds in the motion of a rigid body about a fixed
  point.
\newblock {\em Rev. Roum. Math. Pures et Appl.}, 16(10):1497--1521, 1971.

\bibitem{Kharlamov83}
M.~P. Kharlamov.
\newblock Bifurcation of common levels of first integrals of the {K}ovalevskaya
  problem.
\newblock {\em Prikl. Matem. Mekhan.}, 47(6):737--743, 1983.

\bibitem{KozlovTen96}
V.~V. Kozlov and V.~V. Ten.
\newblock The topology of the accesible region of motion for integrable
  systems.
\newblock {\em Matem. Sbornik}, page (to appear), 1996.
\newblock (in Russian).

\bibitem{Oshemkov91}
A.~A. Oshemkov.
\newblock Fomenko invariants for the main integrable cases of the rigid body
  motion equations.
\newblock In A.~T. Fomenko, editor, {\em Topological Classification of
  Integrable Systems}, volume~6 of {\em Adv. in Soviet Mathematics}, pages
  67--146, Providence, RI, 1991. American Mathematical Society.

\bibitem{Poincare1892}
H.~Poincar\'{e}.
\newblock {\em Les M\'{e}thodes Nouvelles de la M\'{e}canique C\'{e}leste}.
\newblock Gauthier-Villars, Paris, 1892.
\newblock Translated by the AIP, History of Modern Physics and Astronomy,
  Vol.~13, (1993).

\bibitem{SeifertThrelfall80}
H.~Seifert and W.~Threlfall.
\newblock {\em A Textbook of Topology}.
\newblock Academic Press, New York, 1980.

\bibitem{Smale70}
Steven Smale.
\newblock Topology and mechanics {I+II}.
\newblock {\em Inv. Math.}, 10+11:305--331 + 45--64, 1970.

\bibitem{Zieshang94}
R.~St{\"o}cker and H.~Zieshang.
\newblock {\em Algebraische Topologie}.
\newblock Teubner, Stuttgart, 1994.

\bibitem{Tatarinov73}
Ya.~V. Tatarinov.
\newblock On the study of the phase space topology of compact configurations
  with symmetry.
\newblock {\em Vestnik. Mosk. Univ. Math.-Mech.}, 5, 1973.

\bibitem{Zieshang80b}
H.~Zieshang, E.~Vogt, and H.-D. Coldewey.
\newblock {\em Surfaces and Planar Discontinuous Groups}, volume 835 of {\em
  Lect. Notes in Math.}
\newblock Springer, Berlin, 1980.

\end{thebibliography}

\end{document}